\documentclass[useAMS,usenatbib]{mn2e}
\usepackage{aas_macros,graphicx,times,multirow}

\usepackage{graphicx}
\usepackage{standalone}
\usepackage{subfig}
\usepackage{multirow}
\usepackage{array}

\def \rxj{RX\, J1856.5$-$3754}

\title[Optical polarimetry of \rxj]{Evidence for vacuum birefringence from the first optical polarimetry measurement of the isolated neutron star \rxj\thanks{Based on observations collected at the European Organisation for Astronomical Research in the Southern Hemisphere under ESO programme 095.D-0343(A)}}
\author[R. P. Mignani, et al. ]
{\parbox{\textwidth}{R. P. Mignani$^{1,2}$\thanks{E-mail: mignani@iasf-milano.inaf.it},
V. Testa$^{3}$, D. Gonz{\'a}lez Caniulef$^{4}$, R. Taverna$^{5}$, R. Turolla$^{5,4}$,  S. Zane$^{4}$, K. Wu$^{4}$
}
\\ \\
$^{1}$ INAF - Istituto di Astrofisica Spaziale e Fisica Cosmica Milano, via E. Bassini 15, 20133, Milano, Italy\\
$^{2}$ Janusz Gil Institute of Astronomy, University of Zielona G\'ora, Lubuska 2, 65-265, Zielona G\'ora, Poland \\
$^{3}$ INAF - Osservatorio Astronomico di Roma, via Frascati 33, 00040, Monteporzio, Italy \\
$^{4}$ Mullard Space Science Laboratory, University College London, Holmbury St. Mary, Dorking, Surrey, RH5 6NT, UK \\
$^{5}$ Dipartimento di Fisica e Astronomia, Universit\'a di Padova, via Marzolo 8, 35131 Padova, Italy
}

\begin{document}

\date{...}

\pagerange{\pageref{firstpage}--\pageref{lastpage}} \pubyear{...}

\maketitle

\label{firstpage}

\begin{abstract}

The ``Magnificent Seven'' (M7) are a group of radio-quiet Isolated Neutron Stars (INSs) discovered in the soft X-rays through their purely thermal surface emission. Owing to the large inferred magnetic fields ($B\approx 10^{13}$ G), radiation  from these sources is expected to be substantially polarised, independently on the mechanism actually responsible for the thermal emission. A large observed polarisation degree is, however, expected only if quantum-electrodynamics (QED)  polarisation effects are present in the magnetised vacuum around the star. The detection of a strongly linearly polarised signal would therefore provide the first observational evidence of QED effects in the strong-field regime. While polarisation measurements in the soft X-rays are not feasible yet, optical polarisation measurements are within reach also for quite faint targets, like the M7 which have optical counterparts with magnitudes $\approx 26$--$28$. Here, we report on the measurement of optical linear polarisation for the prototype, and brightest member, of the class, \rxj\ ($V\sim 25.5$), the first ever for one of the M7, obtained with the Very Large Telescope. We measured a polarisation degree $\mathrm{P.D.} =16.43\% \pm5.26\%$ and a polarisation position angle $\mathrm{P.A.}=145\fdg39\pm9\fdg44$, computed east of the North  Celestial Meridian. The $\mathrm{P.D.}$ that we derive is large enough to  support the presence of vacuum birefringence, as predicted by QED.
\end{abstract}


\section{Introduction}

A class of seven radio-quiet isolated neutron stars (INSs), a.k.a. the ``Magnificent Seven'' (M7), attracted the interest from the neutron star community soon after their discovery by {\em ROSAT} in the 1990s \cite[see e.g.][for a review]{turolla09}.  Owing to the purely thermal X-ray emission from the star surface ($kT\sim 50$--$100$ eV), unmarred by magnetospheric contributions, the M7 have always been considered unique laboratories to study the neutron star cooling process and hold the promise to provide a measurement of the star radius, which directly bears to the equation of state of matter at supra-nuclear density. The M7 exhibit slow X-ray pulsations ($P\sim 3$--$11$ s), likely produced by an inhomogeneous surface temperature distribution, and relatively large period derivatives ($\dot{P}\sim 10^{-14}$--$10^{-13}$ s s$^{-1}$), from which ages of a few Myrs and dipolar surface magnetic fields $B\sim 10^{13}$--$10^{14}$ G were inferred. Evidence for such strong magnetic fields also came from the discovery of broad ($EW\sim 10$--100~eV) absorption features at few hundred eVs superimposed to the thermal continuum, when interpreted as proton cyclotron and/or bound-free, bound-bound transitions in low-Z atoms. The nearly perfect blackbody (BB) spectrum of the M7 has been a puzzle throughout. Cooling INSs are expected to be covered by an atmosphere, which, however, does not emit a purely BB spectrum \cite[see e.g.][]{pot14}. An alternative
possibility is that the outermost star layers are in a condensed form owing to the large magnetic field and relatively low temperature, giving rise to a so called ``bare'' INS \cite[][]{turolla04}. Spectra from bare INSs would be close to a BB, although with some deviations.

Given the quite strong surface magnetic field of the M7, thermal radiation is expected to be polarised, either if emission is from a bare surface or from an atmosphere \cite[see][]{turolla04,pot14}. The polarisation properties are quite different in the two cases, although there are still uncertainties, especially at optical/ultraviolet (UV) wavelengths. One of the first predictions of quantum-electrodynamics (QED), even before it was properly formulated,
was vacuum birefringence, and, in particular, that a strong magnetic field affects the propagation of light through it \cite[][]{heisen36,weis36}. In thermally emitting INSs, radiation comes from a region comparable with the entire star surface, over which the magnetic field direction changes substantially. In the absence of QED vacuum polarisation effects, this would produce a drastic depolarisation of the radiation collected at infinity \cite[][see also \citealt{taverna15,denis16} and references therein]{Heyletal2003}. Vacuum birefringence dramatically increases the linear polarisation of the observed radiation, from a level of a few $\%$ up to even $\sim 100\%$, depending on the viewing geometry and the surface emission mechanism
\cite[][]{HeylShaviv2000,HeylShaviv2002,Heyletal2003,taverna15,denis16}.
Detecting polarisation in the thermal emission from the surface of an INS will be therefore extremely valuable. First, and independently on the physical conditions of the emitting region, the detection of a large degree of linear polarisation in the signal  would constitute the observational evidence of QED vacuum polarisation effects in the strong-field regime. Second, the polarisation observables can be compared  with emission models and help to uncover the physical conditions of INS surfaces and atmospheres, ideally complementing spectral observations \cite[][]{taverna14,denis16}.

With X-ray polarimetry still moving its first steps and dedicated polarimetric missions, such as {\em XIPE}\footnote{http://www.isdc.unige.ch/xipe}  \cite[][]{soffitta13} and {\em IXPE} \cite[][]{weis13}, having just been proposed, other energy ranges must be explored. Besides the X-rays, all M7 have been detected in the optical/near-UV \cite[][]{kaplan11} where well-tested techniques, even for objects as faint as INSs, can be exploited, as shown by the polarisation measurements in
rotation-powered pulsars (RPPs), for which we refer to Mignani et al.\ (2015) for an exhaustive summary.
Whether the optical/near-UV emission is thermal and arises from the cooling surface is not entirely clear yet. However, in a few sources, notably including the prototype of the class, RX\, J1856.5$-$3754 \cite[][]{walter97}, the low-energy spectral energy distribution (SED) follows quite closely a Rayleigh-Jeans (RJ) distribution (Van Kerkwijk \& Kaplan 2001a). This is a robust indication that optical/near-UV photons are thermal and come from the surface, possibly from a cooler, larger region than that emitting the X-rays \cite[as also supported by {\em XMM-Newton}
data;][]{sart12}.

Owing to their faintness, no optical polarimetric observations have been attempted so far for any of the M7. In this respect, the most promising target is RX\, J1856.5$-$3754. The source is the brightest \cite[$V\sim 25.5$;][]{walter97} and youngest \cite[$0.42$ Myr;][]{Mignani13} among the M7.
Its close distance \cite[$123^{+11}_{-15}$ pc;][]{walter10}, which minimises the effects of foreground polarisation, also concurs to make \rxj\ the
best target.
Here, we present the results of phase-averaged linear polarisation observations of \rxj, the first ever performed for one of the M7.  We have carried out the observations with the Very Large Telescope (VLT) at the ESO Cerro Paranal Observatory (Chile). Observations and data analysis are described in Section \ref{obs}, results are presented in Section \ref{results} and discussed in Section \ref{discuss}. Conclusions follow.

\section{Observations and data reduction}
\label{obs}

\subsection{Observation description}

We measured the optical linear polarisation of \rxj\  using the second Focal Reducer and low dispersion Spectrograph \cite[FORS;][]{app98}, mounted  at the VLT Antu telescope. FORS2 features a red-sensitive mosaic MIT detector (two $4\mathrm k\times2\mathrm k$ CCDs aligned along the X axis)
and is equipped with polarisation optics to measure either linear or circular time-averaged polarisation. The polarisation optics consist of a Wollaston prism as a beam splitting analyser and two super-achromatic phase  retarder 3$\times$3 plate  mosaics (Boffin 2014).  These retarder plates are installed on rotatable mountings to be moved in and out of the light path. In the  imaging polarimetry mode (IPOL), a strip mask is produced in the FORS2 focal plane to separate the extraordinary and ordinary beams of polarised light as this passes through the retarder plate. The strip is formed by lacing every second Multi-Object Spectrograph (MOS) slit jaw carrier arm across the instrument field of view. In this way, by taking two frames displaced by $22^{\prime\prime}$ in the Y direction,  images corresponding to the extraordinary and ordinary beams are recorded on two adjacent MOS slitlets. 

We used the four standard IPOL  half-wave retarder plate angles of $0^\circ$, $22\fdg5$,
$45^\circ$, and $67\fdg5$, which correspond to the retarder plate orientation relative  to the Wollaston prism and are usually set with an accuracy better than 0\fdg1\footnote{ www.eso.org/sci/facilities/paranal/instruments/fors.html}. Both the axis of the detector optics and the zero point of the half
wave retarder plate angle  are set such that the polarisation position angle is measured eastward from the North Celestial Meridian.

The observations of \rxj\ were executed in service mode in May and June 2015. We used the FORS2 low gain, normal readout (200 Kpix/s), standard-resolution mode (0\farcs25/pixel; $2 \times 2$ binning), and the high efficiency v$_{\rm HIGH}$ filter ($\lambda=555.0$ nm, $\Delta\lambda=61.6$ nm).   In order to
cope with the variable sky polarisation background, each observation block (OB) incorporated exposures ($\sim$ 720 s) for each of the four retarder plate angles.  A total of  11 OBs were executed, for a total exposure of 7920 s per angle. Observations were performed  in dark time, with average seeing of 0\farcs84, airmass below 1.1, and clear sky conditions.

\subsection{Data reduction}

A number of short exposures ($< 3$ s) of polarised
standard stars \cite[][]{mat71,whi92} were acquired for calibration purposes, as part of the FORS2 calibration plan. Observations of unpolarised standard stars were also acquired  to monitor the FORS2 instrumental polarisation at the CHIP1 aim position. We applied the same reduction procedure (bias-subtraction and flat-fielding) for both \rxj\ and the standard stars using standard routines in {\sc IRAF}\footnote{IRAF is distributed by the National Optical Astronomy Observatories, which are operated by the Association of Universities for Research in Astronomy, Inc., under cooperative agreement with the National Science Foundation.}. Twilight flat-field images with no retarder plate along the light path were acquired on the same nights as the science images. For each OB, we finally obtained four reduced images one for each of the four retarder  plate angles. We computed the FORS2 astrometry from the short exposure (60 s) science acquisition image using stars from the Guide Star  Catalogue 2 \cite[GSC-2;][]{las08} as a reference. After accounting for all uncertainties (star centroids, GSC-2 absolute coordinate accuracy, etc.) our astrometric solution turned out to be accurate to better than 0\farcs1 rms.   We updated the \rxj\ position at the epoch of our FORS2 observations (2015.37) by correcting for its proper motion $\mu$ using its reference coordinates (epoch 1999.7) measured with the {\em Hubble Space Telescope} \cite[][]{kap02}. The position uncertainty associated with the propagation of the proper motion errors ($\pm 1$ mas yr$^{-1}$) is negligible. The \rxj\ coordinates at the epoch of our VLT observations are, then:  $\alpha =18^{\rm h}  56^{\rm m} 36\fs03$; $\delta  = -37^\circ 54\arcmin 37\farcs11$ (J2000) and we used them as a reference to identify it in our VLT images.

\begin{figure}
\includegraphics[width=8.5cm]{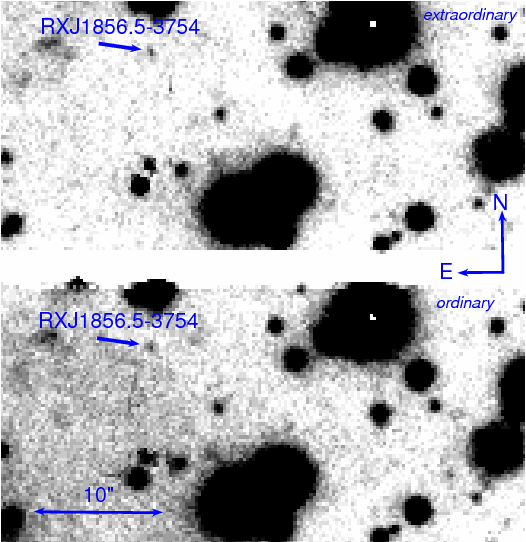}
\caption{Section of the FORS2 IPOL v$_{\rm HIGH}$-band image of the \rxj\ field obtained from the co-addition of all exposures taken with all retarder plate angles.  The optical counterpart to \rxj\ is marked by the arrow and labelled.  The image is divided in two parts, corresponding to two adjacent MOS slitlets, as the effect of the FORS2 polarisation optics, which split the incoming light into two beams. The top and bottom part of the image correspond to the extraordinary and ordinary beams, respectively, whereas the white stripe corresponds to the gap between the two adjacent MOS slitlets (see Sectn. 2.1 for details).    \label{fc}}
\end{figure}

\subsection{Polarisation measurement}

Following the FORS2 manual \cite[][]{bof14}, the degree of linear polarisation of a source is calculated from the normalised Stokes parameters $P_{\mathrm U} \equiv U/I$ and $P_{\mathrm Q} \equiv Q/I$:

\begin{equation}
P_{\mathrm Q } = \frac{1}{2} \ \left \{ {\left(\frac{f^\mathrm o -
f^\mathrm e}{f^\mathrm o + f^\mathrm e}
\right)}_{\alpha=0^{\circ}} -
\left(\frac{f^\mathrm o - f^\mathrm e}{f^\mathrm o + f^\mathrm e} \right)_{\alpha=45^{\circ}}\right \}\\
\end{equation}

\begin{equation}
P_{\mathrm U} = \frac{1}{2} \ \left \{ {\left(\frac{f^\mathrm o -
f^\mathrm e}{f^\mathrm o + f^\mathrm e}
\right)}_{\alpha=22.5^{\circ}} -
\left(\frac{f^\mathrm o - f^\mathrm e}{f^\mathrm o + f^\mathrm e} \right)_{\alpha=67.5^{\circ}}\right \}\\
\end{equation}

where $f^\mathrm o$ and $f^\mathrm e$ are the source fluxes in the ordinary (o) and extraordinary (e) beams, respectively, for each of the four retarder plate angles $\alpha$  ($0^{\circ}$, $22\fdg5$, $45^{\circ}$ and $67\fdg5$).
 The linear polarisation degree ($\mathrm{P.D.}$) and position angle ($\mathrm{P.A.}$) are, then,
determined from $P_{\mathrm U}$ and $P_{\mathrm Q}$ as follows:

\begin{equation}
\label{pd}
\mathrm{P.D.} = \sqrt{{P_\mathrm Q}^{2} + {P_\mathrm U}^{2}}\\
\end{equation}

\begin{equation}
\label{poang}
\mathrm{P.A.} = \frac{1}{2} \ \arctan\left(\frac{P_\mathrm U}{P_\mathrm Q} \right).\\
\end{equation}

By convention, in the following, we report the value of $\mathrm{P.D.}$ in per cent rather than in unity fractions. We note that for IPOL observations in the V band, the zero angle of the half wave  retarder plate is offset from 0\degr\ by 1\fdg8 \cite[][]{bof14} and this term must be subtracted from the $\mathrm{P.A.}$ defined in equation \ref{poang}. 

In order to increase the signal-to-noise ratio, we co-added all the reduced science  images of \rxj\ taken with the same retarder plate angle ($0^\circ$, $22\fdg5$,
$45^\circ$, and $67\fdg5$) and used them to compute $f^\mathrm o$ and $f^\mathrm e$, hence the  linear  polarisation degree.  For each angle, we aligned the single images using the IRAF tasks {\tt ccdmap} and {\tt ccdtrans}, with an average accuracy of a few hundredths of a pixel.  We then applied the image co-addition with the IRAF routine {\tt combine} and filtered cosmic ray hits  using an average $\sigma$ clipping. Figure \ref{fc} shows a section of the image obtained from the co-addition of all exposures taken with all retarder plate angles, zoomed on the \rxj\ position.

To obtain a more robust detection of our target and improve on the photometry, we adopted a technique regularly used for studies of crowded 
stellar fields. First of all, we carefully aligned and registered all the images, regardless of the retarder plate angle, and co-added them to create a master image that we used as a reference for the source detection in each of the  co-added $0^\circ$, $22\fdg5$, $45^\circ$, $67\fdg5$ retarder plate angle images. This method, is well-suited to detect faint objects and improve the precision on the object centroid determination.  As a subsequent step, we used the task {\em daofind} in the {\sc IRAF}  package {\tt daophot} \cite[][]{ste87,ste94}, which we ran on the master image to obtain a master source list that we used as a reference for the object detection in each of the four images above by keeping the objects' centroids  fixed.  
In this way, we minimise the degrees of freedom in the photometry fit and ensure a better flux measurement for fainter objects, decreasing the photometry error by $\sim$20\%.
From the master source list we also selected a number of stars to compute the model PSF for each of the four co-added images above by fitting the intensity profiles  of a number of suitable stars in the field of view.  

Finally, we used the \rxj\ position in the master source list and the computed model PSF as input to the  {\em ALLSTAR} routine in {\tt daophot} to measure its flux through PSF photometry in each of the co-added $0^\circ$, $22\fdg5$, $45^\circ$, $67\fdg5$ retarder plate angle images, and in both the ordinary and extraordinary beams. According to our experience with polarisation measurements of such faint targets \cite[see, e.g.][]{Mignani15}, PSF photometry is more robust than aperture photometry.  In particular, we fitted the model PSF to the \rxj\ intensity profile within an area of 10 pixel radius (2\farcs5), estimated from the growth curves of the reference stars and measured the sky background in an annulus of inner radius $10.5$ pixels and width $10$ pixels (2\farcs6 and 2\farcs5), respectively. As a last step, we  corrected the \rxj\ flux for the
atmospheric extinction using, for each image, the average of the measured airmass values and the extinction coefficients in the v$_{\rm HIGH}$ filter computed by 
the FORS2  quality control team\footnote{www.eso.org/qc.}.

We applied the same procedure as above to the polarised standard stars to calibrate the \rxj\ polarisation. From the comparison between the measured and reference values for the standard stars, we found that our polarimetry has, on average, an absolute uncertainty of only $0.13\% \pm 0.06\%$
in $\mathrm{P.D.}$ 
and of  0\fdg2$\pm$0\fdg4
in $\mathrm{P.A.}$.
These values  are consistent with both the trending analysis carried out by the FORS2 quality control team and with the independent analysis of Cikota et al.\ (2016), based upon six years of FORS2 observations of polarimetry standard stars. At the same time,  from the observations of the unpolarised standard stars we found that there is no evidence of significant instrumental polarisation ($0.09\%\pm0.06\%$) at the CHIP1 aim position. 

\section{Results}
\label{results}

\subsection{The  \rxj\ polarisation}

For the photometry parameters defined above, we measured a $\mathrm{P.D.}=16.43\% \pm5.26\%$  and  a  $\mathrm{P.A.}=145\fdg39\pm9\fdg44$ for \rxj, where we computed the errors as described in  Fossati et al.\ (2007). 
The target is faint, and hence the uncertainties are dominated by statistics rather than calibration.

With the measurement presented here, there are now six isolated neutron stars for which   the phase-averaged (or phase-resolved) linear polarisation has been measured. Five of them are RPPs: the Crab and Vela pulsars, PSR\ B0540$-$69, PSR\, B1509$-$58, and PSR\, B0656+14 (see, Mignani et al.\ 2015 and references therein), i.e. the five brightest RPPs identified in the optical (Mignani 2011). Therefore, \rxj\ is the first isolated neutron star other than a RPP for which a measurement  of the optical polarisation has been obtained. An upper limit on the optical linear polarisation for the magnetar 4U\, 0142+61 (5.6\% at the 90\% c.l.) has been recently reported by Wang et al.\ (2015)\footnote{One measurement  and two upper limits on the infrared polarisation have been obtained for other three magnetars (Israel, G. L., p.c.)}. 

We note that the polarisation level measured for \rxj\  seems to be above what measured for RPPs, for which the $\mathrm{P.D.}$  is usually below 10\% (see, Mignani et al. \ 2015).  The preferential orientation of the polarisation vector is also different. Previous optical and/or radio polarisation measurements in RPPs point toward a substantial alignment between the polarisation vector and the proper motion direction \cite[][and references therein]{Mignani15}.  
In the optical, this alignment has been measured for the Crab and Vela pulsars (Moran et al.\ 2013; 2014) and, possibly, PSR\, B0656+14 (Mignani et al.\ 2015), whereas nothing can be said for both PSR\, B0540$-$69 and PSR\, B1509$-$58 for which no proper motion has been measured yet.
 In the case of \rxj, the polarisation $\mathrm{P.A.}$ that we measured with the VLT ($\rm 145\fdg39\pm9\fdg44$)  and that of the proper motion ($\mathrm{P.A.}^\mu =100\fdg2\pm0\fdg2$; Kaplan et al.\ 2002) differ significantly (by $\sim 45^\circ$),  even accounting for the substantial uncertainty of the former measurement,  
so the source does not appear to fit in this picture. 
It should be noted, however, that the emission process which produces the polarised optical radiation is different for RPPs and thermally emitting INSs, such as RX\, J1856.5$-$3754. In young RPPs, such as the Crab and Vela pulsars,  the optical emission is produced in the neutron star magnetosphere (Pacini \& Salvati 1983), i.e. along a preferred direction, whereas in thermally emitting INSs the optical radiation comes from the entire cooling neutron star surface.

\subsection{Remarks on the possibility of polarisation contamination}
\label{contam}

\rxj\ lies $\sim 1\fdg4$ away from the centre of the dusty molecular cloud complex in the Corona Australis region (CrA). The distance to the CrA complex is, however, uncertain (see Neuh\"auser \& Forbrich  2008 for a discussion).  Based upon measurements of the interstellar reddening from the Hipparcos and Tycho catalogues, Knude \& H{\o}g (1998) determined a  distance of  $\sim 170$ pc.  On the other hand,  based upon the orbit solution of the eclipsing double-lined spectroscopic binary TY CrA (Casey et al.\ 1998) a distance as low as $\sim 130$ pc was inferred.  Therefore, we cannot  firmly rule out either that the CrA dark cloud complex is foreground to \rxj\  \cite[$123^{+11}_{-15}$ pc;][]{walter10} or that this is embedded in it.  In this case, although \rxj\ is clearly away from the densest regions of the CrA complex, as seen from the Digitised Sky Survey images,  it would still be possible that  the more tenuous cloudsÕ limbs extend to the neutron star position. This might represent a possible source of contamination to our polarisation measurement, which we would have to take into account. Indeed, stars embedded in, or in the background of, dust clouds  are known to exhibit a non-zero degree of polarisation, produced by starlight scattering onto the dust grains \cite[e.g.][for a review]{draine04}. The observed degree of linear polarisation $p(\lambda)$ usually follows the empirical Serkowski's law \cite[][]{serk73}:

\begin{equation}
p(\lambda)=p_\mathrm{max}\exp[-K\ln^2(\lambda/\lambda_\mathrm{max})]
\end{equation}

where $\lambda$ is the radiation wavelength; polarisation attains a maximum,  $p(\lambda)=p_\mathrm{max}$,  at $\lambda=\lambda_\mathrm{max}$. In general $K$ linearly depends on $\lambda_\mathrm{max}$ \cite[e.g.][]{wilk82,whi92,cik16}, while the maximum polarisation $p_\mathrm{max}$ is found to be a function of the reddening \cite[][and references therein]{serk73}, $p_\mathrm{max}\sim 0.09 E(B-V)$. Since $\lambda_\mathrm{max}\sim 555\ \mathrm{nm}$ \cite[][]{serk75}, which coincides with the pivot wavelength of the v$_{\rm HIGH}$ filter (Sectn. 2.1),
the effects of dust on our measurement could be potentially relevant. 
The measured interstellar extinction towards \rxj\ is $A_\mathrm V\sim 0.12$ \cite[][]{vkk01a}, accounting for both gas and dust contributions, which gives $E(B-V)\sim 0.04$. Therefore, assuming that the CrA complex is, indeed, foreground to \rxj, the  maximum degree of polarisation predicted by the Serkowski's law in the v$_{\rm HIGH}$ filter ($\lambda\simeq\lambda_\mathrm{max}$) would be $p_\mathrm{max}\sim0.36\%$. 

The capacity of dust grains to produce polarisation upon scattering depends on their degree of alignment, i.e. on how much their spin angular momenta  point in the same direction \cite[a still very debated issue; see e.g.][]{draine04}. 
By adopting, as we did, $p_\mathrm{max}= 0.09 E(B-V)$, some degree of alignment of the grains is already accounted for. However, it has been found that, along some directions, higher values of $p_\mathrm{max}$ are required to match the data with the Serkowski's law in the B and V bands \cite[e.g.][]{clay92}. 
This seems to imply a higher alignment of the grains along those lines-of-sight, which is associated with a reddening typically much larger than that observed along the line-of-sight to  \rxj.
Nonetheless, even assuming that a thin wisp of material with highly aligned grains lies in between us and \rxj\ 
\cite[e.g., connected with the bow-shock nebula associated with the source;
][]{vkk01b},
 it seems very unlikely that this could contribute to the observed polarisation degree by as much as  $1\%$. 
 
\begin{figure}
\includegraphics[width=8.5cm]{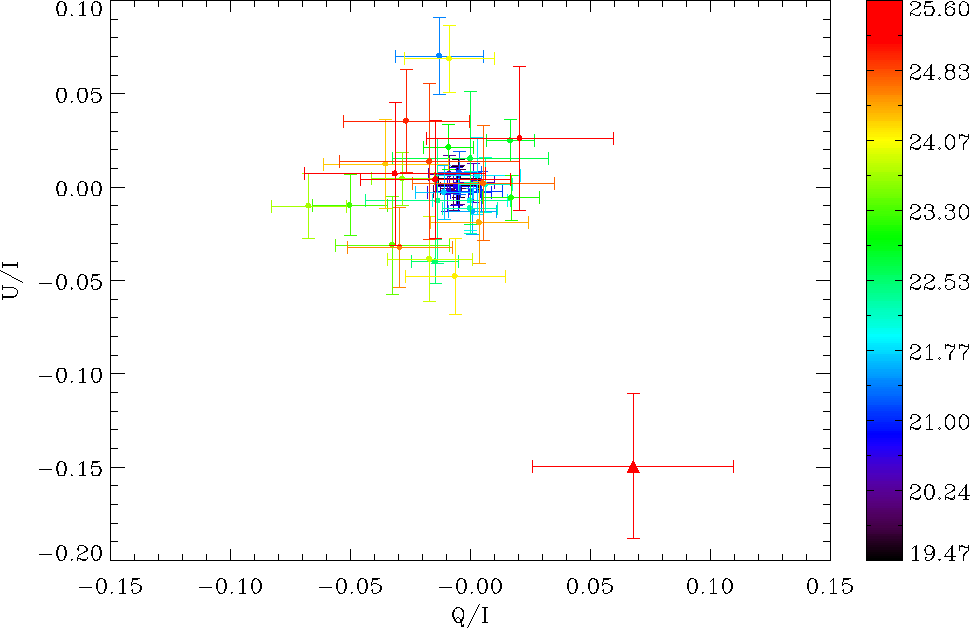}
\caption{Plot of the normalised Stokes parameters $U/I$ and $Q/I$ for a selected sample of 42 field stars (filled point; top left) and for \rxj\ (filled triangle, bottom right). We warn that the values of $U/I$ and $Q/I$ plotted here, computed from Eqn. 1 and 2, are in unity fractions and so would be the value of $\mathrm{P.D.}$ computed from them (Eqn. \ref{pd}). The colour scale on the right corresponds to the stars' V-band magnitude. 
\label{polstars}}
\end{figure}

This conclusion is confirmed by the values of the Stokes parameters $P_\mathrm Q$ and $P_\mathrm U$ measured for a set of 42 stars randomly selected within a region of a few arcmin size in the \rxj\ field, as shown in Figure \ref{polstars}.  We selected  stars in the magnitude range $\sim$20--26 and whose positions are more than 1\arcmin\ away from the CCD edges, to avoid the effects of off-axis instrumental polarisation. We filtered out obvious outliers through a $3\sigma$ clipping and stars with 
errors larger than $\sim5\%$ on the $\mathrm{P.D.}$. Such stars are either saturated or with positions too close to the edges of the MOS slitlets, which affects the measurement of the fluxes $f^\mathrm o$ and $f^\mathrm e$.
We note that for the field stars the intrinsic degree of polarisation is very small or even statistically compatible with zero, hence, the observed polarisation would be essentially that produced by dust. On the other hand, for \rxj\ the measured degree of polarisation is substantial ($\ga 11\%$), ruling out that it could be due to dust. These arguments indicate that, in our case, dust contamination is also quite unimportant quantitatively. 

Another possible source of polarisation contamination could be the \rxj\ bow-shock nebula itself.
Indeed, free electrons in the shocked interstellar medium (ISM) may modify the polarisation signal.  However, the estimated number density ($\la 3\ \mathrm{cm}^{-3}$) in the shocked ISM surrounding RX\, J1856.5$-$3754 \citep{vkk01b} is too small to make propagation effects such as scattering or Faraday conversion relevant in this context.

 \section{Discussion}
\label{discuss}

\subsection{Comparison with theoretical models}
\label{model}

The polarisation properties of radiation emitted by the cooling surface of INSs  are strongly influenced by QED vacuum polarisation effects as photons propagates in the strongly magnetised vacuum that surrounds the star \cite[e.g.][]{HeylShaviv2000,HeylShaviv2002,Heyletal2003, vanALai}. A systematic analysis of the polarisation properties ($\mathrm{P.D.}$ and $\mathrm{P.A.}$) and of their dependence on the INS physical and geometrical parameters was presented by \cite{taverna15} in the case of pure BB surface emission.
A specific application to the optical and X-ray polarisation properties of  RX\, J1856.5$-$3754 was recently discussed by \cite{denis16}, who also considered more realistic surface emission models, either a gaseous atmosphere (GA) or a condensed surface (CS), properly accounting for general-relativistic ray-bending and vacuum polarisation effects. Their analysis assumed a dipolar magnetic field with an intensity at the pole
$B_\mathrm p= 10^{13} \ \rm G$, as inferred from the spin-down measurements by \cite{vkk08}, and a dipole-induced surface temperature distribution with $T^{\infty}_\mathrm p=60$~eV at the pole, truncated at $T^{\infty}_{\mathrm e}=40$~eV at the equator, as it follows from the most recent spectral study of \citet[][see \citealt{denis16} for more details and \citealt{popov16} for a discussion of alternative thermal maps]{sart12}, where the temperatures are measured with respect to an observer at infinity.
The star mass and radius were assumed to be $M_{\rm NS}=1.5M_\odot$ and $R_{\rm NS}=12 \ {\rm{km}}$, respectively, which imply a gravitational red-shift factor at the surface $1+z=1.26$. In particular, \cite{denis16} calculated the phase-averaged $\mathrm{P.D.}$ and $\mathrm{P.A.}$ as a function of the angles $\chi$
and $\xi$, which correspond to the inclination of the line-of-sight (LOS) and of the magnetic axis with respect to the INS spin axis, respectively, both in the soft X-rays ($0.12$--$0.39$ keV) and in the Cousins B filter ($\lambda=445 \ \rm{nm}$, $\Delta\lambda=94 \ \rm{nm}$).

\begin{figure*}
\includegraphics[width=16.cm]{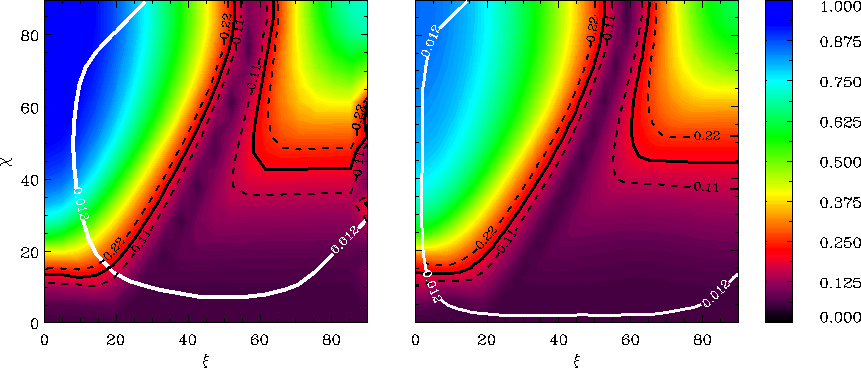}
\caption{Contour plot of the phase-averaged linear polarisation degree in the VLT v$_{\rm HIGH}$ band for the isotropic blackbody (left) andthe gaseous atmosphere (right) models.  The thick white line marks the locus in the $\xi$--$\chi$ plane where the computed pulsed fraction matches the observed value, $1.2\%$ \protect\cite[][]{tienmer07}. The solid black line corresponds to the measured VLT optical polarisation of RX\, J1856.5$-$3754 while the dashed lines correspond to the $\pm 1 \sigma$ error ($\mathrm{P.D.}=16.43 \pm 5.26\%$). \label{bb+atmo}}
\end{figure*}

\begin{figure*}
\includegraphics[width=16.cm]{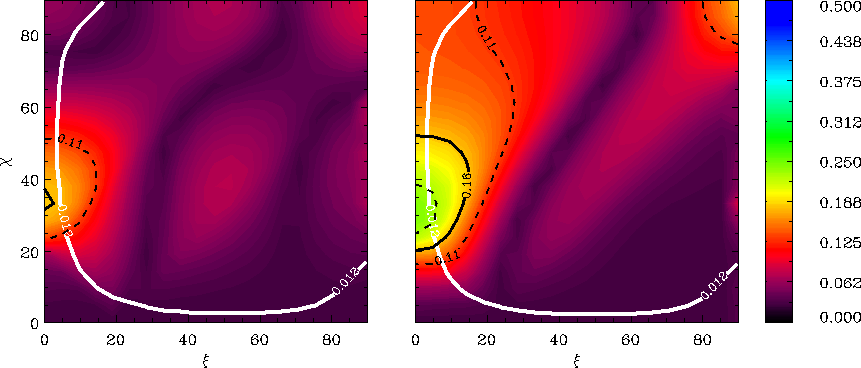}
\caption{Same as in Figure \ref{bb+atmo} for emission from a condensed surface in the free (left) and fixed ion limit (right). 
\label{fix+free} }
\end{figure*}

Using the same approach described in \cite{denis16}, we computed the  polarisation properties in the v$_{\rm HIGH}$ band ($\lambda=555.0$ nm, $\Delta\lambda=61.6$ nm), in which our data were collected. We carried out the calculations for three different surface emission models: (i) an isotropic BB, assuming that the radiation is $100\%$ polarised in the extraordinary mode, (ii) a magnetised, completely ionised hydrogen atmosphere, and (iii) a CS model \cite[both in the fixed and free ion limit; see again][for details]{denis16}. Results are summarised in Figures \ref{bb+atmo} and \ref{fix+free}, which show the observed $\mathrm{P.D.}$ as a function of $\chi$ and $\xi$, and in Table \ref{angles}. For the BB and the GA emission models there is a range of angles that is consistent with the constraints set by the value of the X-ray pulsed  fraction, \cite[$1.2~\%$, ][]{tienmer07}, and the observed
$\mathrm{P.D.}$, $16.43\% \pm 5.26\%$. In the first case, the angle between the LOS and the spin axis
is   $\chi =15\fdg6^{+1.8}_{-1.7}$ and the angle between the magnetic axis and the  spin axis is $\xi = 18\fdg0^{+2.4}_{-1.8}$, while in the second case it is $\chi = 14\fdg0^{+2.3}_{-3.0}$ and  $\xi = 3\fdg1^{+0.8}_{-0.4}$.  CS models provide a solution only in the fixed ion limit,
$\chi = 21\fdg7^{+5.9}_{-4.4}$ and $\xi = 5\fdg5^{+1.4}_{-1.3}$ , together with $\chi = 51\fdg9^{+37.3}_{-14.7}$ and $\xi = 2\fdg9^{+13.2}_{-0.4}$; the central value of the measured $\mathrm{P.D.}$ is never recovered in the free ion case. For all the considered emission models,
our analysis points at small values of $\xi$, and $\chi \sim 15^\circ$--$20^\circ$ (in the case of  CS emission model and the fixed ions limit there is a second solution at $\chi\sim50^\circ$). This narrows the (broader) constraints set on the two geometrical angles by \cite{ho07}, $\chi\approx 20^{\circ}$--$45^\circ$ and $\xi\la 6^{\circ}$ (or vice versa), on the basis of their pulse profile fits with atmospheric models. With the present estimate of the geometrical angles, a putative radio beam from \rxj\ would not intersect the LOS, either assuming the pulsars average beaming factor, $f=0.1$, or using the expression by \cite{tauris98} to relate $f$ to the spin period \cite[see also][]{ho07}.

\begin{table}
\begin{center}
\caption{Summary of the values of $\chi$ and $\xi$ obtained from the comparison of the measured polarisation degree and different emission models (first column).}
\label{angles}
\begin{tabular}{lll} \hline
model  & $\chi (^\circ)$  &  $\xi (^\circ)$   \\  \hline
 BB      & $15.6^{+1.8}_{-1.7}$  & $18.0^{+2.4}_{-1.8}$     \\
      &  &    \\
 GA     & $14.0^{+2.3}_{-3.0}$ &  $3.1^{+0.8}_{-0.4}$    \\
      &  &    \\
 CS$^\dagger$      & $21.7^{+5.9}_{-4.4}$   & $5.5^{+1.4}_{-1.3}$     \\
       &  &    \\
                 & $51.9^{+37.3}_{-14.7}$ & $2.9^{+13.2}_{-0.4}$ \\ \hline
                       &  &    \\
\cite{ho07} &     $\approx$ 20--45      &  $\la 6$ \\
      &  &    \\
\hline
 \end{tabular}
\end{center}
 $^\dagger$ Fixed ion limit.
\end{table}

\subsection{Test of QED predictions}
\label{qed}

\begin{figure*}
\includegraphics[width=16.cm]{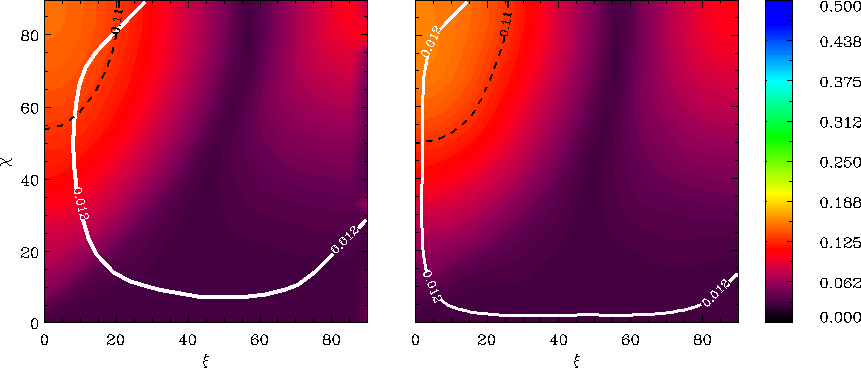}
\caption{Same as in Figure \ref{bb+atmo}  but without accounting for vacuum polarisation effects. \label{BB+atmo-off} }
\end{figure*}

\begin{figure*}
\includegraphics[width=16.cm]{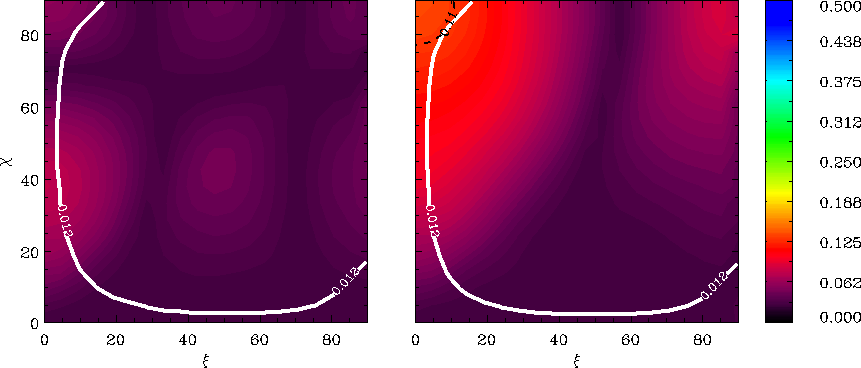}
\caption{Same as in Figure \ref{fix+free} but without accounting for vacuum polarisation effects. \label{fix+free-off} }
\end{figure*}

Polarimetric observations offer an unique opportunity to experimentally verify the predictions of QED vacuum polarisation effects in the strong field limit. To test this, we  recomputed the optical $\mathrm{P.D.}$ for RX\, J1856.5$-$3754 neglecting vacuum polarisation effects. Results are shown in Figure \ref{BB+atmo-off} and \ref{fix+free-off}. In all cases, the computed phase-averaged $\mathrm{P.D.}$ is substantially reduced, because the depolarisation induced by the frame rotation is much
more effective at small values of the adiabatic radius \cite[with no QED vacuum polarisation effects included the adiabatic radius coincides with the star radius;][]{HeylShaviv2002,vanALai,taverna15,denis16}. The minimum attainable phase-averaged $\mathrm{P.D.}$ is still within the measured range for the BB, GA, and CS (fixed ions) emission models, while it is not for the CS model in the free ion limit. However, the most likely measured value is never reproduced by models with no QED vacuum polarisation effects and the highest attainable $\mathrm{P.D.}$, which is anyway below the measurement, requires a very particular geometry of the source, i.e. an aligned rotator ($\xi \approx 0^{\circ}$), seen equator-on ($\chi \approx 90^{\circ}$). The observed high value of the $\mathrm{P.D.}$ is therefore a strong indicator for the presence of vacuum polarisation effects around RX\, J1856.5$-$3754.

\begin{figure*}
\includegraphics[width=16.cm]{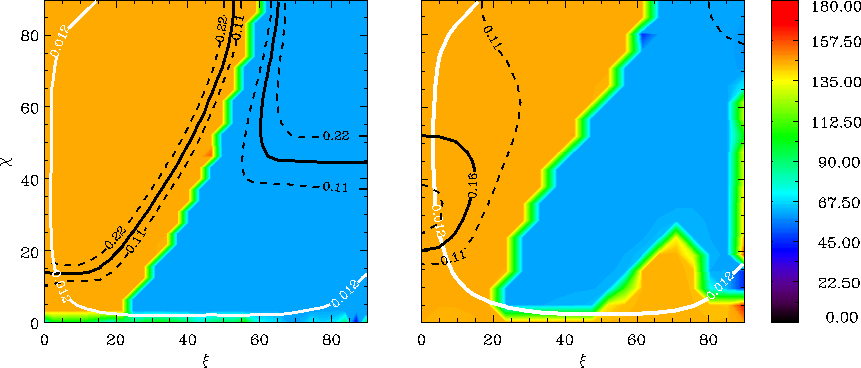}
\caption{Contour plot of the phase-averaged linear polarisation position angle in the VLT v$_{\rm HIGH}$ band for emission from a gaseous atmosphere (left) and a
condensed surface (fixed ion limit; right). In each case a rotation by an angle $\beta$ around the LOS has been applied (see
Sectn. \ref{geom} for details). The labelled contours have the same meaning as in Figure \ref{bb+atmo}. \label{polang}}.
\end{figure*}

\subsection{Constraints on neutron star viewing orientations}
\label{geom}

The polarisation position angle $\mathrm{P.A.}$ provides independent information about the viewing orientations of the INS. We remark, however, that care must be taken in comparing the observed value with those derived from simulations since the  $\mathrm{P.A.}$ depends on the choice of a reference direction in the plane of the sky. \cite{denis16} computed the phase-averaged $\mathrm{P.A.}$
assuming as reference direction
the projection of the star spin axis on the plane of the sky. In the case of the GA emission model, they found that the phase-averaged $\mathrm{P.A.}$ was nearly constant over almost the entire region where $\chi> \xi$ and assumes a value of $\sim 90^{\circ}$. This implies that, when ``phase-averaging'' the signal, the direction of the electric field of the observed radiation should oscillate mostly in a direction perpendicular to the star spin axis projected on the sky. For the CS emission model (here we consider only the fixed ion limit since it matches the $\mathrm{P.D.}$ constraint), the situation is similar but somehow reversed, with $\mathrm{P.A.} \sim 0^\circ$ in the region where $\chi > \xi$, meaning that the electric field oscillates mainly along the projection of the spin axis.

Our VLT  polarimetry observations of RX\, J1856.5$-$3754 give a polarisation position angle $\mathrm{P.A.}=145\fdg39\pm9\fdg44$, where this angle is computed using as reference direction the  North Celestial Meridian. As for the $\mathrm{P.D.}$, we recomputed the $\mathrm{P.A.}$ in the VLT v$_{\rm HIGH}$ band, obtaining results very
similar to those of \cite{denis16}, i.e.
$\mathrm{P.A.}=90^\circ$ and $0^\circ$ for the GA and the CS model, respectively.
As discussed above, we need to introduce a rotation by an angle $\beta$ around the LOS, to account for the different reference
direction in the simulations
and in the observations\footnote{This amounts to a rotation of the Stokes parameters by
an  angle $2\beta$ around the LOS, which clearly leaves the $\mathrm{P.D.}$
unchanged.}. In this case, $\beta$ obviously corresponds to angle between  the projection of the neutron star spin axis on the plane of the sky and the  North Celestial Meridian. Therefore, we can obtain for the first time the absolute direction of the spin-axis of  RX\, J1856.5$-$3754 in the plane of the sky. We found that, for the atmosphere model, a rotation angle $\beta=55\fdg39 \pm 9\fdg44$ (or $235\fdg39 \pm 9\fdg44$) reproduces the observed value, for the viewing angles derived in section \ref{model} ($\chi = 14\fdg0^{+2.3}_{-3.0}$, $\xi =3\fdg1^{+0.8}_{-0.4}$). Similarly, a rotation angle $\beta=145\fdg39 \pm 9\fdg44$ (or $325\fdg39 \pm 9\fdg44$) is required for the CS model (fixed ions), see Figure \ref{polang}.

From the value of $\beta$, we can derive the angle $\theta$ between the projection of the neutron star spin axis on the plane of the sky and the neutron star proper motion vector, 
where  $\theta = \mathrm{P.A.}^\mu-\beta$ and  $\mathrm{P.A.}^\mu=100\fdg2 \pm 0\fdg2$ \cite[][]{kap02}.
For the atmospheric case ($\beta=55\fdg39 \pm 9\fdg44$), 
we obtain 
that the direction of the spin-axis of  RX\, J1856.5$-$3754 
is tilted  westward of the
 neutron star proper motion vector 
  by an angle $\theta =44\fdg81\pm9\fdg64$. In the case of the CS
emission model ($\beta=145\fdg39 \pm 9\fdg44$), instead,
we obtain 
that the projection of the spin axis is tilted eastward of the proper motion vector by $\theta =45\fdg19\pm9\fdg64$. Therefore, in both cases the projection of the spin axis and the proper motion vector are significantly misaligned.

Although this is still a debated issue, it has been suggested that in the Crab and Vela pulsars the direction of the spin axis projected in the plane of the sky, assumed to coincide with the axis of symmetry of their X-ray pulsar wind nebulae (PWNe), coincides with that of the proper motion \cite[][see also \citealt{kaplan08}]{helfand01,pavlov01}. \cite{Noutsos2012}
presented a statistical analysis based on a large sample of pulsars with proper motion and polarisation measurement, and found evidence for  pulsar spin-velocity alignment, although  the possibility of orthogonal spin-velocity configurations could not be excluded. Such an alignment derives from that of the pulsar rotation axis and the velocity vector. The latter can be produced if the proto-neutron star kick resulted from multiple thrusts over a time longer than the star initial period.
On the other hand, no correlation is expected if the acceleration phase is shorter \cite[][]{spruit98}. Therefore, the $\sim 45^\circ$ misalignment that we found in \rxj\ might not be peculiar, also accounting for the fact that the velocity vector might have changed over the source lifetime, $\sim 0.42$ Myr, as the result of the interaction with the Galactic gravitational potential  (e.g., Mignani et al.\ 2013).  
For comparison, in the case of the radio-quiet RPP Geminga, of age ($\approx 0.3$ Myr) comparable to \rxj, the pulsar spin axis (again assumed to coincide with the axis of symmetry of its PWN) and the proper motion vector are aligned (Pavlov et al.\ 2010). This would suggest that their relative orientation has  not been affected by the pulsar orbit in the Galactic potential (Pellizza et al.\ 2005).

\section{Summary and conclusions}

We presented the first measurement of optical polarisation from a thermally emitting isolated neutron star, RX\, J1856.5$-$3754. We measured a linear polarisation degree $\mathrm{P.D.}=16.43\% \pm5.26\%$ and a polarisation position angle $\mathrm{P.A.}=145\fdg39\pm9\fdg44$.  At variance with the case of RPPs, in which polarisation measurements give an indication of the magnetic field direction in the magnetosphere (Mignani et al.\ 2015 and references therein), in the case at hand the polarisation observables are directly linked to the properties of the surface (or atmospheric) layers of the neutron star. As already pointed out by Gonzalez Caniulef et al.\ (2016), optical measurements 
complemented by measurements of X--ray polarisation
can be used to discriminate between the case of a gaseous atmosphere and a condensed neutron star surface layer for different viewing geometries.
Thus, our measurement paves the way to future X-ray polarimetry measurements in the soft X rays (e.g., Marshall et al.\ 2015).
In any case,  we found that, independently on how
thermal photons are produced, such a high value of  linear polarisation in the signal is extremely unlikely to be reproduced by models in which QED vacuum polarisation  effects are not accounted for.
Therefore, our VLT polarisation measurement constitutes the very first observational support for the predictions of QED vacuum polarisation effects. 
Follow-up optical polarimetry observations of \rxj\ will confirm and improve our measurement and make the observational support to the QED predictions more robust. Optical polarimetry measurements of some of the other M7 would be obviously  important to consolidate our results, although such measurements would be more challenging owing to the fact that the other M7 are even fainter than \rxj\ (Kaplan et al.\ 2011). Imaging polarimetry capabilities at the next generation of 30-40m class telescopes, such as the European Extremely Large Telescope (E-ELT), would then play a crucial role in testing QED predictions of vacuum polarisation effects on a larger sample.
Measurements of the circular optical polarisation, never obtained so far for any isolated neutron star other than the Crab pulsar (Wiktorowicz et al.\ 2015),  should also be pursued. Circular polarisation for thermal radiation coming from the neutron surface is zero in the approximation adopted in the present work (a sharp boundary for the adiabatic region; see Taverna et al.\ 2015) and, according to our numerical simulations, it is expected not to exceed $1$--$2\%$ even accounting for the presence of an intermediate region, the only place where circular polarisation can be created \cite[][]{taverna14,taverna15}. The measurement of such a small polarisation degree could be attainable, again, only exploiting  the high throughput of the next generation of extra-large telescopes.

\section*{Acknowledgments}
We thank the anonymous referee for his/her very constructive comments to our manuscript. RPM acknowledges financial support from the project TECHE.it. CRA 1.05.06.04.01 cap 1.05.08 for the project "Studio multi-lunghezze
d'onda da stelle di neutroni con particolare riguardo alla emissione di altissima energia". DGC acknowledges the financial support by CONICYT through a
'Becas Chile' fellowship No. 72150555.

\label{lastpage}


\begin{thebibliography}{99}

\bibitem[\protect\citeauthoryear{Appenzeller et al.}{1998}]{app98}  Appenzeller I., Fricke K., F\"urtig W., et al., 1998, The Messenger, 94, 1

\bibitem[\protect\citeauthoryear{Boffin}{2014}]{bof14} Boffin H.M.J., 2014, The FORS2 User Manual

\bibitem[\protect\citeauthoryear{Casey et al.}{1998}]{cas98}Casey B.W., Mathieu R. D., Vaz L. P. R., Andersen J., \& Suntzeff N. B., 1998, AJ, 115, 1617

\bibitem[\protect\citeauthoryear{Cikota et al.}{2016}]{cik16}Cikota A., Patat F., Cikota S., Faran T., 2016, MNRAS, in press, arXiv:1610.00722

\bibitem[\protect\citeauthoryear{Clayton et al.}{1992}]{clay92} Clayton G.C. et al., 1992, \apj, 385, L53

\bibitem[\protect\citeauthoryear{Draine}{2004}]{draine04} Draine B.T., 2004, in Saas-Fee Advanced Course ``The Cold Universe'', eds. D. Pfnenniger, Y. Revaz, vol. 32.  Springer-Verlag, Berlin Heideiberg, p. 231


\bibitem[\protect\citeauthoryear{Fossati et al.}{2009}]{Fossati} Fossati L., Bagnulo S., Mason E., Landi Degl'Innocenti E., 2007, Proc. of "The Future of Photometric, Spectroscopic, and Polarimetric Standardization", eds. C. Sterken, ASP Conference Series

\bibitem[Gonz{\'a}lez Caniulef et al.(2016)]{denis16} Gonz{\'a}lez Caniulef D., Zane  S., Taverna R., Turolla R., \& Wu K., 2016, MNRAS, 459, 3585

\bibitem[\protect\citeauthoryear{Heisenberg \& Euler}{1936}]{heisen36} Heisenberg W., Euler H. 1936, Z. Physik, 98, 714

\bibitem[\protect\citeauthoryear{Helfand, Gotthelf \& Halpern}{2001}]{helfand01} Helfand D.J., Gotthelf E.V., Halpern, J.P. 2001, ApJ, 556, 380

\bibitem[\protect\citeauthoryear{Heyl \& Shaviv}{2000}]{HeylShaviv2000} Heyl J.S., Shaviv N.J., 2000, MNRAS, 311, 555

\bibitem[\protect\citeauthoryear{Heyl \& Shaviv}{2002}]{HeylShaviv2002}Heyl J.S., Shaviv N.J., 2002, Phys. Rev. D, 66, 023002

\bibitem[\protect\citeauthoryear{Heyl, Shaviv \& Lloyd}{2003}]{Heyletal2003}Heyl J.S., Shaviv N.J., Lloyd D., 2003, MNRAS, 342, 134

\bibitem[Ho(2007)]{ho07} Ho, W.~C.~G.\ 2007, \mnras, 380, 71

\bibitem[\protect\citeauthoryear{Kaplan et al.}{2002}]{kap02} Kaplan, D. L., van Kerkwijk, M. H., Anderson, J., 2002, ApJ, 571, 447

\bibitem[\protect\citeauthoryear{Kaplan et al.}{2008}]{kaplan08}Kaplan D.L., Chatterjee S., Gaensler B.M., Anderson
J. 2008, ApJ, 677, 1201

\bibitem[\protect\citeauthoryear{Kaplan et al.}{2011}]{kaplan11} Kaplan D. L., Kamble  A., van Kerkwijk M. H., Ho W. C. G.,  2011, ApJ, 736, 117


\bibitem[\protect\citeauthoryear{Knude \& H{\o}g}{2011}]{knu98} Knude J., H{\o}g E., 1998, A\&A, 338, 897 

\bibitem[\protect\citeauthoryear{Lasker et al.}{2008}]{las08} Lasker B. M., Lattanzi M. G., McLean B. J., et al., 2008, AJ, 136, 735

\bibitem[\protect\citeauthoryear{Mathewson \& Ford}{1971}]{mat71} Mathewson D. S. \& Ford V. L., 1971, MNRAS, 153, 525

\bibitem[\protect\citeauthoryear{Mignani}{2011}]{}  Mignani R.P., 2011, ASpR, 47, 1281

\bibitem[\protect\citeauthoryear{Mignani et al.}{2013}]{Mignani13} Mignani R. P., et al., 2013, MNRAS, 429, 3517

\bibitem[\protect\citeauthoryear{Mignani et al.}{2015}]{Mignani15} Mignani R. P., Moran P., Shearer A., Testa, V., S\l{}owikowska A., Rudak B., Krzeszowski K., Kanbach G.,  2015, MNRAS, 583, 105

\bibitem[\protect\citeauthoryear{Marshall et al.}{2015}]{Mar15} Marshall H. L., Schulz N. S., Windt D. L., Gullikson E. M., Craft M., Blake E., Ross C., 2015, Proceedings of the SPIE, Vol. 9603

\bibitem[\protect\citeauthoryear{Moran et al.}{2013}]{Moran13} Moran P., Shearer  A., Mignani R.P., S\l{}owikowska A., De Luca A., Gouiffes C., Laurent P., 2013, MNRAS, 433, 2564

\bibitem[\protect\citeauthoryear{Moran et al.}{2014}]{Moran14} Moran P., Mignani R.P., Shearer  A., 2014, MNRAS, 445, 835


\bibitem[\protect\citeauthoryear{Neuh\"auser \& Forbrich }{2008}]{neu08} 	Neuh\"auser R., Forbrich J., 2008, Handbook of Star Forming Regions, Volume II: The Southern Sky ASP Monograph Publications, Vol. 5, Bo Reipurth eds, p.735
	
\bibitem[\protect\citeauthoryear{Noutsos et al.}{2012}]{Noutsos2012} Noutsos A., Kramer M., Carr P., Johnston S.  2012, MNRAS, 423, 2736

 \bibitem[\protect\citeauthoryear{Pacini \& Salvati}{1983}]{ps83} Pacini F. \& Salvati M., 1983, ApJ, 274, 369, 1983


\bibitem[\protect\citeauthoryear{Pavlov et al.}{2001}]{pavlov01}  Pavlov G. G., Kargaltsev O. Y., Sanwal D., Garmire G. P.,  2001, ApJ 554, L189

\bibitem[\protect\citeauthoryear{Pavlov et al.}{2010}]{pavlov10} Pavlov G. G., Bhattacharyya S., Zavlin V. E. 2010, ApJ, 715, 66

\bibitem[\protect\citeauthoryear{Pellizza et al.}{2005}]{pel05}Pellizza L. J., Mignani R. P., Grenier I. A., Mirabel I. F., 2005, A\&A, 435, 625



\bibitem[\protect\citeauthoryear{Popov et al.}{2016}]{popov16} Popov S.B., Taverna R., Turolla R., 2016, \mnras, in press (arXiv:1610.05050)

\bibitem[\protect\citeauthoryear{Potekhin}{2014}]{pot14} Potekhin, A.~Y.\ 2014, Physics Uspekhi, 57, 735 (arXiv:1403.00774)

\bibitem[\protect\citeauthoryear{Sartore et al.}{2012}]{sart12} Sartore N., Tiengo A., Mereghetti S., De Luca A., Turolla R., Haberl F., 2012, A\&A, 541, 66

\bibitem[\protect\citeauthoryear{Serkowski}{1973}]{serk73} Serkowski K., 1973, in the Proceedings of Interstellar Dust and Related Topics. IAU Symposium no. 52, eds. J. M. Greenberg and H. C. van de Hulst. Reidel, Dordrecht Boston, p.145

\bibitem[\protect\citeauthoryear{Serkowski et al.}{1975}]{serk75} Serkowski K., Mathewson D.S., Ford W.L. 1975, \apj, 196 261


\bibitem[\protect\citeauthoryear{Soffitta et al.}{2013}]{soffitta13} Soffitta P., et al., 2013, Experimental Astronomy, 36, 523

\bibitem[\protect\citeauthoryear{Spruit \&
Phinney}{1998}]{spruit98} Spruit H., Phinney E.S. 1998, Nature,
393, 139

\bibitem[\protect\citeauthoryear{Stetson}{1987}]{ste87} Stetson P.B., 1987, PASP, 99, 191 

\bibitem[\protect\citeauthoryear{Stetson}{1994}]{ste94} Stetson P.B., 1994, PASP, 106, 250 

\bibitem[\protect\citeauthoryear{Tauris \& Manchester}{1998}]{tauris98} Tauris T.M., Manchester R.N. 1998, MNRAS, 298, 62

\bibitem[\protect\citeauthoryear{Taverna et al.}{2014}]{taverna14} Taverna R., Muleri F., Turolla R., Soffitta P., Fabiani S., Nobili L., 2014, MNRAS, 438,1686

\bibitem[Taverna et al.(2015)]{taverna15} Taverna, R., Turolla, R., Gonzalez Caniulef, D., Zane, S.\ 2015, \mnras, 454, 3254

\bibitem[Tiengo \& Mereghetti(2007)]{tienmer07} Tiengo, A., \& Mereghetti, S.\ 2007, \apjl, 657, L101

\bibitem[\protect\citeauthoryear{Turolla et al.}{2004}]{turolla04} Turolla R.,  Zane S., Drake J.J.,  2004, ApJ, 603, 265

\bibitem[\protect\citeauthoryear{Turolla}{2009}]{turolla09} Turolla R., 2009, Astrophysics and Space Science Library, 357, 141

\bibitem[\protect\citeauthoryear{van Adelsberg and Lai}{2006}]{vanALai} van Adelsberg M. \& Lai D., 2006, MNRAS, 373, 1495

\bibitem[\protect\citeauthoryear{van Adelsberg \& Perna}{2009}]{vanAdelsbergPerna}
van Adelsberg, M., Perna, R., 2009, MNRAS, 399, 1523

\bibitem[van Kerkwijk \& Kaplan(2008)]{vkk08} van Kerkwijk M.~H., \& Kaplan D.~L.\ 2008, \apjl, 673, L163

\bibitem[\protect\citeauthoryear{van Kerkwijk \& Kulkarni}{2001a}]{vkk01a} van Kerkwijk M.H., Kulkarni S.R. 2001a, \aap, 378, 986

\bibitem[\protect\citeauthoryear{van Kerkwijk \& Kaplan}{2001b}]{vkk01b} van Kerkwijk M.H., Kulkarni S.R. 2001b, \aap, 380, 221

\bibitem[\protect\citeauthoryear{Walter \& Matthews}{1997}]{walter97} Walter F. M.,  \&  Matthews L. D., 1997, Nature, 389, 358

\bibitem[\protect\citeauthoryear{Walter et al.}{2010}]{walter10}  Walter F. M., Eisenbeiss T.,  Lattimer J. M., Kim B., Hambaryan V., Neuh\"auser R., 2010, ApJ, 724, 669

\bibitem[\protect\citeauthoryear{Wang et al.}{2015}]{} Wang Z., Tanaka Y. T., Wang C., Kawabata K. S., Fukazawa Y., Itoh R., Tziamtzis A., 2015, ApJ, 814, 89

\bibitem[\protect\citeauthoryear{Weisskopf}{1936}]{weis36} Weisskopf V. 1936, Kongelige Danske Videnskaberns Selskab, Mathematisk-Fysisky Maddeleser 14, 1

\bibitem[\protect\citeauthoryear{Weisskopf et al.}{2013}]{weis13} Weisskopf M.C. et al. 2013, Proceedings
of the SPIE, 8859, 885908

\bibitem[\protect\citeauthoryear{Whittet et al.}{1992}]{whi92} Whittet D. C. B.,  Martin P. G., Hough J. H., Rouse M. F., Bailey J. A., Axon D. J., 1992, ApJ, 386, 562

\bibitem[\protect\citeauthoryear{Wiktorowicz et al.}{2015}]{wik15}  Wiktorowicz S.,  Ramirez-Ruiz E., Illing R. M. E., Nofi L., 2015, American Astronomical Society, AAS Meeting \#225

\bibitem[\protect\citeauthoryear{Wilking et al.}{1982}]{wilk82} Wilking B.A., Lebofsky M.J., Rieke G.H. 1982, \aj, 87, 695



\end{thebibliography}
\end{document}